\documentclass[submission,copyright,creativecommons,11pt]{eptcs}
\providecommand{\event}{EXPRESS 2011}

\usepackage{amsmath,amssymb,latexsym}
\usepackage{graphicx}

\usepackage{breakurl}

\newtheorem{theorem}{Theorem}[section]
\newtheorem{definition}[theorem]{Definition}

\newtheorem{proposition}[theorem]{Proposition}
\newtheorem{example}[theorem]{Example}

\def\altbox{\hspace{2mm}\nolinebreak\null\nolinebreak\hfill\Box}

\def\name#1{\mbox{\sc #1}}
\def\sos#1#2{{\def\arraystretch{1.6}\begin{array}{c}#1\\\hline%
#2\end{array}}}

\newcommand{\Nil}{{\sf nil}}
\newcommand{\ignore}[1]{}

\newcommand{\gor}{\;\big|\;}
\newcommand{\rec}{{\sf rec} \:}
\newcommand{\bisimilar}{\ensuremath{\sim}}
\newcommand{\fase}{\texttt{FASE}}

\newcommand{\nar}[2]{\xrightarrow{#1}_{#2}}

\newcommand{\rnar}[2]{\stackrel{#1}{\rightsquigarrow}_{#2}}
\newcommand{\onar}[2]{\stackrel{#1}{\mapsto}_{#2}}

\newcommand{\B}{\mathbb{A}} 
\newcommand{\Bt}{\B_{\tau}} 
\newcommand{\X}{{\cal X}}
\newcommand{\Pg}{\tilde{\PG}}
\newcommand{\PG}{{\mathbb P}}
\newcommand{\Sg}{\tilde{\SG}}
\newcommand{\SG}{{\mathbb S}}
\newcommand{\synsort}[1]{{\cal L}(#1)}

\newcommand{\ur}{{\cal U}}
\newcommand{\rop}{\triangleright}

\newcommand{\denote}[1]{[\![ #1 ]\!]}
\newcommand{\denoter}[1]{[\![ #1 ]\!]_r}
\newcommand{\denotes}[1]{[\![ #1 ]\!]_s}

\newcommand{\clean}{{\sf clean}}
\newcommand{\unlab}{{\sf unmark}}

\newcommand{\NF}{\PG_{\it rn}}
\newcommand{\Nf}{\Pg_{\it rn}}

\title{Read Operators and their Expressiveness in Process Algebras%
\thanks{This work was supported by the PRIN Project `Paco:
Performability-Aware Computing: Logics, Models, and Languages}}
\author{Flavio Corradini \quad Maria Rita Di Berardini
\institute{School of Science and Technology,
Computer Science Division,\\ University of Camerino}
\email{flavio.corradini@unicam.it,
mariarita.diberardini@unicam.it}
\and Walter Vogler
\institute{Institut f\"ur Informatik,\\
Universit\"at Augsburg, Germany}
\email{vogler@informatik.uni-augsburg.de}
}
\def\titlerunning{Read Operators and their Expressiveness in Process
Algebras}
\def\authorrunning{F. Corradini, M.R. Di Berardini \& W. Vogler}
\begin{document}
\maketitle

\begin{abstract}
We study two different ways to enhance PAFAS, a process algebra for
modelling asynchronous timed concurrent systems, with non-blocking
reading actions. We first add reading in the form of a read-action
prefix operator. This operator is very flexible, but its somewhat
complex semantics requires two types of transition relations. We also
present a read-set prefix operator with a simpler semantics, but with
syntactic restrictions. We discuss the expressiveness of read
prefixes; in particular, we compare them to read-arcs in Petri nets
and justify the simple semantics of the second variant by showing that
its processes can be translated into processes of the first with
timed-bisimilar behaviour. It is still an open problem whether the
first algebra is more expressive than the second; we give a number of
laws that are interesting in their own right, and can help to find a
backward translation.
\end{abstract}

\section{Introduction}\label{intro}

Non-blocking reading is an important feature e.g.\ for proving the liveness
of MUTEX solutions under the progress assumption (aka weak fairness). We
study the first process algebra with non-blocking read actions, where \lq
read\rq\ refers to accessing a variable, e.g.\ modelled as a separate
process $Var$. Observe that read is an activity of $Var$, and in a setting
with explicit modelling of data, it would rather be an output than an input
action of $Var$.

Non-blocking reading is known from Petri nets, where it has been added in
the form of read arcs; these allow multiple concurrent reading of the same
resource, a quite frequent situation in many distributed systems. Read arcs
represent {\em positive context conditions}, i.e.\ elements which are
needed for an event to occur, but are not affected by it. As argued
in~\cite{MR95}, the importance of such elements is twofold.  Firstly, they
allow a faithful representation of systems where the notion of ``reading 
without consuming'' is commonly used, like database systems~\cite{R94} or
any computation framework based on shared memory. Secondly, they allow to
specify directly and naturally a level of concurrency greater than in
classical nets: two transitions reading the same place may also occur
simultaneously; in classical nets, the transitions would be connected to
the place by loops  (namely, i.e.\ reading is modelled
through a ``rewrite'' operation) which does not allow the simultaneous
execution of two tasks that read the same resource. 
Read arcs have been used
to model a variety of applications such as transaction serialisability
in
databases~\cite{R94}, concurrent constraint programming~\cite{MR94},
asynchronous systems~\cite{VOGLER02}, and cryptographic
protocols~\cite{CW01}. Reading is also related to the notion of {\em
persistence} e.g.\ in several calculi for describing and analysing security
protocols; in particular, persistent messages (that can be read but not
consumed) are used to model that every message can be remembered by the spy
(see~\cite{CCA+07} and the references therein).

Semantics and expressivity of read arcs have been studied e.g.\ in the
following:~\cite{CH93} discusses a step semantics;~\cite{BHR06} shows that
timed Petri nets with read arcs unify timed Petri nets and timed automata.
Finally, \cite{VOGLER02} shows that read arcs add relevant expressivity;
the MUTEX problem can be solved with nets having read arcs but not with
ordinary nets having no read arcs.

In this paper, we present two different ways to enhance PAFAS~\cite{CVJ02},
a process algebra for modelling asynchronous timed concurrent systems, with
non-blocking reading actions. PAFAS was introduced for evaluating the
worst-case efficiency of asynchronous systems. It was also used in
\cite{CDV06,CDV06-2} for studying (weak) fairness of actions and components
in system computations, similarly to results of~\cite{VOGLER02} for a Petri
net setting. This fairness requires that an action has to be performed (a
component has to perform an action, resp.), whenever it is enabled
continuously in a run. Fairness can be defined in an intuitive but
complicated way in the spirit of \cite{CS84,CS87}, and we proved that each
everlasting (or non-Zeno maximal) timed run is fair and vice
versa~\cite{CDV06}. We used these characterisations in \cite{CDV06-2} to
prove that Dekker's MUTEX algorithm satisfies the respective liveness
property under the assumption of {\em fairness of components}, while this
fails under {\em fairness of actions}. To improve this, one needs suitable
assumptions about the hardware, cf.~\cite{Ray86}, namely that reading a
value from a storage cell is non-blocking; to model this we introduce
specific reading prefixes for PAFAS.

We first add reading in the form of a read-action prefix $\alpha \rop Q$
(the new process language is called PAFAS$_r$), which behaves as $Q$ but,
like a variable or a more complex data structure, can also be read with the
action $\alpha$. Since being read should not change the state, $\alpha$ can
be repeated until the execution of some ordinary action of $Q$. Thus, e.g.\
$a \rop b.\Nil$ can perform any number of $a$'s until it terminates via an
ordinary $b$. The operational semantics for $\alpha \rop Q$ needs two types
of transition relations to properly deal e.g.\ with sequences of read
actions.

Under some syntactic restrictions, the semantics can be simplified. To be
still able to express sequences of read actions directly, we introduced a
read-set operator $\{a_1, \cdots, a_n\} \rop Q$ in the language PAFAS$_s$.
In~\cite{CDV09}, we already used PAFAS$_s$ to show the correctness of
Dekker's algorithm: regarding some actions as reading, this algorithm
satisfies MUTEX liveness already under the assumption of {\em fairness of
actions}. It had long been an open problem how to achieve such a result in
a process algebra~\cite{Wal89}. The simpler semantics of PAFAS$_s$ is
helpful for building tools. Indeed, we have already proved some MUTEX
algorithms correct or incorrect with the aid of the automated verification
tool \fase~\cite{BCC+11}. We plan to continue this work by also considering
the efficiency of MUTEX algorithms and other systems.

In this paper, we study PAFAS$_r$ and PAFAS$_s$ further with  special
attention to expressiveness. The first issue is that PAFAS$_r$ models
non-blocking reading in an intuitive way, while the necessary restrictions
in case of PAFAS$_s$ are not so obvious. In fact, the investigations for
this paper have disclosed that the restrictions in~\cite{CDV09} still
allowed processes with a contra-intuitive semantics. To rectify this subtle
mistake, we give an improved definition of \emph{proper} PAFAS$_s$
processes\footnote{Luckily, the model of Dekker's algorithm in~\cite{CDV09}
is also proper as defined here.}, and we show how to translate each proper
process $Q$ into a PAFAS$_r$ process whose timed behaviour is bisimilar and
even isomorphic to that of $Q$. This shows at the same time that a proper
process really has an intuitive behaviour and that PAFAS$_r$ is at least as
expressive as the proper fragment of PAFAS$_s$.

In this paper, we additionally show that safe Petri nets with read-arcs as
in~\cite{VOGLER02} can be modelled with proper PAFAS$_s$ processes. It is
still an open problem whether PAFAS$_r$ is more expressive than PAFAS$_s$;
we present a number of laws that are interesting in their own right and
give a backward translation for a fragment of PAFAS$_r$. Constructing a
general backward translation seems to be related to finding an expansion
law for PAFAS$_r$ processes, a law that is not even known for standard
PAFAS processes.

We have also extended the correspondence between fair and everlasting
runs; thus, also in PAFAS$_r$ and in PAFAS$_s$, we capture fairness
with timed behaviour. To demonstrate the extended expressiveness of
reading with a concrete example, we prove that no finite-state process
in standard PAFAS has the same fair language as $a \rop b.\Nil$
(Theorem~\ref{theorem-expr-read-actions}).

The rest of the paper is organised as follows. Sections~\ref{PAFAS-r}
and~\ref{PAFAS-s} introduce PAFAS$_r$ and PAFAS$_s$ with their respective
timed operational semantics and prove result regarding $a \rop b$. 
Section~\ref{secmapping} provides a mapping from PAFAS$_s$ to PAFAS$_r$ and
presents the result for Petri nets. The backward translation is discussed
in Section~\ref{sec-backward-mapping}. Finally,
Section~\ref{sec-conclusion} presents some concluding remarks. Some proofs
can be found in the appendices.

\section{A process algebra for describing read behaviours} \label{PAFAS-r}
In this section, we introduce PAFAS$_r$ and  give a first expressiveness
result. PAFAS is a CCS-like process description language \cite{Mil89} (with
a {\em TCSP}-like parallel composition~\cite{BHR84}), where actions are
atomic and instantaneous but have associated an upper time bound (either
$0$ or $1$, for simplicity) interpreted as a maximal time delay for their
execution. As explained in \cite{CVJ02}, these upper time bounds can be
used for evaluating the performance of asynchronous systems, but do not
influence \emph{functionality} (which actions are performed); so compared
to CCS, also PAFAS treats the full functionality of asynchronous systems.
W.r.t. the original language, here we introduce the new {\em read prefix}
$\rop$ to represent non-blocking behaviour of processes. Intuitively, the
term $\alpha\rop P$ models a process like a variable or a more complex data
structure that behaves as $P$ but can additionally be read with $\alpha$:
since being read does not change the state, $\alpha$ can be performed
repeatedly until the execution of some ordinary action of $P$, and it does
not block a synchronisation partner (a reading process) as described below.

We use the following notation. $\B$ is an infinite set of {\em visible
actions}. An additional action $\tau$ is used to represent internal
activity, which is unobservable for other components. We define
$\Bt=\B\cup\{\tau\}$. Elements of $\B$ are denoted by $a,b,c,\dots$ and
those of $\Bt$ by $\alpha,\beta,\dots$. Actions in $\Bt$ can let time $1$
pass before their execution, i.e.\ 1 is their maximal delay. After that
time, they become \emph{urgent} actions written $\underline{a}$ or
$\underline{\tau}$; these cannot be delayed. The set of urgent actions is
denoted by $\underline{\B}_\tau = \{\underline{a} \:|\: a \in \B \}\cup \{
\underline{\tau} \}$ and is ranged over by $\underline{\alpha},
\underline{\beta}, \dots$. Elements of $\Bt \cup \underline{\B}_\tau$ are
ranged over by $\mu$. $\X$ (ranged over by $x, y, z, \ldots$) is the set of
process variables, used for recursive definitions. $\Phi \: : \: \Bt
\rightarrow \Bt$ is a {\it general relabelling function} if the set $\{
\alpha \in \Bt \: | \: \emptyset \neq \Phi ^{-1}(\alpha) \neq \{ \alpha
\}\}$ is finite and $\Phi(\tau)= \tau$. Such a function can also be used to
define {\em hiding}: $P/A$, where the actions in $A$ are made internal, is
the same as $P[\Phi_A]$, where the relabelling function $\Phi_A$ is defined
by $\Phi_A(\alpha)= \tau$ if $\alpha \in A$ and $\Phi_A(\alpha)=\alpha$ if
$\alpha \notin A$.

We assume that time elapses in a discrete way\footnote{PAFAS is not time
domain dependent, meaning that the choice of discrete or continuous time
makes no difference for the testing-based semantics of asynchronous
systems, see~\cite{CVJ02} for more details.}. Thus, an action prefixed
process $a.P$ can either do action $a$ and become process $P$ (as usual in
CCS) or can let one time step pass and become $\underline{a}.P$;
$\underline{a}$ is called {\em urgent} $a$, and $\underline{a}.P$ cannot
delay $a$, but \emph{as a stand-alone process} can only do $a$ to become
$P$. In the following, initial processes are just processes of a standard
process algebra extended with $\rop$. General processes include all
processes reachable from the initial ones according to the operational
semantics to be defined below.

The sets $\Pg_1$ of {\em initial (timed) process terms} $P$ and $\Pg$ of
(general) {\em (timed) process terms} $Q$ is generated by the following
grammar:

\begin{eqnarray*}
P&::=& \Nil \gor x \gor \alpha. P \gor \alpha \rop P \gor P+P \gor P
\:\|_A \: P \gor P[\Phi] \gor \rec x.P\\
Q &::=& P \gor \underline{\alpha} .P \gor \mu\triangleright Q \gor Q+Q
\gor Q \:\|_A\: Q \gor Q[\Phi] \gor \rec x.Q
\end{eqnarray*}

where $\Nil$ is a constant, $x \in \X$, $\alpha \in \Bt$, $\mu \in \Bt \cup
\underline{\B}_\tau$, $\Phi$ is a general relabelling function and $A
\subseteq \B$ possibly infinite. We say that a variable $x \in \X$ is {\em
guarded} in $Q$ if it only appears in the scope of some $\mu \in \Bt \cup
\underline{\B}_\tau$. We assume that recursion is {\em guarded}, i.e.\ for
$\rec x.Q$ variable $x$ is guarded in $Q$. A process term is {\em closed}
if every variable $x$ is bound by the corresponding $\rec x$-operator; the
set of closed timed process terms in $\Pg$ and $\Pg_1$, simply called {\em
processes} and {\em initial processes} resp., is denoted by $\PG$ and
$\PG_1$ resp.

We briefly describe the operators. The $\Nil$-process cannot perform any
action, but may let time pass without limit. A trailing $\Nil$ will often
be omitted, so e.g.\ $a.b+c$ abbreviates $a.b.\Nil + c.\Nil$. $\mu.Q$ is
(action-)prefixing known from CCS. Read-prefixed terms $\alpha \rop Q$ and
$\underline{\alpha} \rop Q$ behave like $Q$ except for the (lazy and
urgent, resp.) non-blocking action $\alpha$. In both cases $\alpha$ is
always enabled until component $Q$ evolves via some ordinary action;
moreover, $\underline{\alpha}$ stays urgent even if it is performed.
$Q_1+Q_2$ models the choice between processes $Q_1$ and $Q_2$. $Q_1
\,\|_A\, Q_2$ is the parallel composition of two processes $Q_1$ and $Q_2$
that run in parallel and have to synchronise on all actions from $A$; this
synchronisation discipline is inspired from {\em TCSP}. $Q[\Phi]$ behaves
as $Q$ but with the actions changed according to $\Phi$. $\rec x.Q$ models
a recursive definition. We often use equations to define recursive
processes, e.g.\ $P \Leftarrow a.P + b$; in contrast, $\equiv$ stands for
syntactically equal. Below we use the (syntactic) sort of a process that
contains all visible actions the process can ever perform.

\begin{definition}\rm\label{def-syntactic-sort}({\sl sort})
For a general relabelling function $\Phi$ let $ib(\Phi) = \{a \in \B \,|\,
\emptyset \neq \Phi^{-1}(a) \neq \{a\}\}$ (the image base of $\Phi$); by
definition of a general relabelling function, $ib(\Phi)$ is finite. The
{\em sort} of $Q \in \Pg$ is the set $\synsort{Q} = \{a \in \B \,|\, a
\mbox{ occurs in } Q\} \cup \bigcup_{\Phi \mbox{ occurs in } Q} \,
ib(\Phi)$.
\end{definition}

The transitional semantics describing the functional behaviour of PAFAS$_r$
terms indicates which actions they can perform. We need two different
transition relations $\onar{\alpha}{}$ and $\rnar{\alpha}{}$ to describe,
resp., the ordinary and the reading behaviour of PAFAS$_r$ processes. The
functional behaviour is the union of these two kinds of behaviour.

\begin{definition}\label{PACTSOS}\rm({\sl functional operational
semantics})
Let $Q\in \Pg$ and $\alpha \in \Bt$. We say that $Q\nar{\alpha}{} Q'$
if $Q\onar{\alpha}{} Q'$ or $Q\rnar{\alpha}{} Q'$, where the SOS-rules
defining the transition relations $\onar{\alpha}{} \subseteq (\Pg
\times \Pg)$ (the {\em ordinary action transitions}) and
$\rnar{\alpha}{} \subseteq (\Pg \times \Pg)$ (the {\em read action
transitions}) for $\alpha \in \Bt$, are given in Tables
\ref{OrdBehaviour} and \ref{ReadBehaviour}, resp.\footnote{We do here
without functions $\clean$ and $\unlab$, used e.g.\ in~\cite{CDV06} to
get a
closer relationship between states of untimed fair runs and timed
non-Zeno runs. They do not change the behaviour (up to an injective
bisimulation) and would complicate the setting.}. As usual, we write
$Q \nar{\alpha}{} Q'$ if $(Q,Q')\in\nar{\alpha}{}$ and
$Q\nar{\alpha}{}$ if  $Q \nar{\alpha}{} Q'$ for some $Q' \in \Pg$; and
analogously for other types of transition relations.
\end{definition}

\begin{table}[tbh]
\small
\[
\begin{array}{c}
\name{Pref$_o$}\; \sos{\mu\in\{\alpha, \underline{\alpha}\}}
{\mu.P \onar{\alpha}{}P}\qquad \quad

\name{Read$_o$}\; \sos{Q \onar{\alpha}{} Q'}{\mu \rop Q
\onar{\alpha}{} Q'}\qquad \quad

\name{Sum$_{o}$}\; \sos{Q_1\onar{\alpha}{}Q'}{Q_1+Q_2
\onar{\alpha}{} Q'}\\

\name{Par$_{o1}$} \sos{\alpha\notin A,\;Q_1\onar{\alpha}{}Q'_1}
{Q_1\|_A Q_2 \onar{\alpha}{} Q'_1\|_A Q_2 }\quad \quad

\name{Par$_{o2}$} \sos{\alpha\in A,\;Q_1\onar{\alpha}{}Q'_1,\;
Q_2\nar{\alpha}{}Q'_2}{Q_1\|_A Q_2 \onar{\alpha}{} Q'_1\|_A Q'_2}\\

\name{Rel$_{o}$} \sos{Q\onar{\alpha}{}Q'}
{Q[\Phi]\onar{\Phi(\alpha)}{}Q'[\Phi]}\quad \quad

\name{Rec$_{o}$} \sos{Q \{\rec x.Q/x\} \onar{\alpha}{} Q'}
{\rec x.Q \onar{\alpha}{} Q'}\\
\end{array}\]
\caption{Ordinary behaviour of PAFAS$_r$ processes}
\label{OrdBehaviour}
\end{table}
Rule \name{Pref}$_o$ in Table~\ref{OrdBehaviour} describes the behaviour of
an action-prefixed process as usual in CCS. Note that timing can be
disregarded: when an action is performed, one cannot see whether it was
urgent or not, and thus $\underline{\alpha}.P \onar{\alpha}{}P$;
furthermore, $\alpha.P$ has to act {\em within} time 1, i.e.\ it can also
act immediately, giving $\alpha.P \onar{\alpha}{}P$. Rule \name{Read$_o$}
says that $\mu \rop Q$ performs the same ordinary actions as $Q$ removing
the read-prefix at the same time. Note that in rule \name{Par$_{o_2}$}, an
ordinary action transition can synchronise with both an ordinary and a
\emph{read action} transition. The other rules are as expected. Symmetric
rules have been omitted.

\begin{table}[tbh]
\small
\[
\begin{array}{c}
\name{Read$_{r1}$}\; \sos{\mu\in\{\alpha,
\underline{\alpha}\}} {\mu \rop Q \rnar{\alpha}{} \mu \rop Q}\quad
\quad
\name{Read$_{r2}$}\; \sos{Q \rnar{\alpha}{} Q'}{\mu \rop Q
\rnar{\alpha}{}
\mu \rop Q'}\quad
\name{Sum$_{r}$}\; \sos{Q_1\rnar{\alpha}{}Q'_1}{Q_1 + Q_2
\rnar{\alpha}{}
Q'_1 + Q_2}\\
\name{Par$_{r1}$} \sos{\alpha\notin A,\;Q_1\rnar{\alpha}{}Q'_1}
{Q_1\|_A Q_2 \rnar{\alpha}{}Q'_1\|_A Q_2}\quad \quad
\name{Par$_{r2}$} \sos{\alpha\in A,\;Q_1\rnar{\alpha}{}Q'_1,\;
Q_2\rnar{\alpha}{}Q'_2}{Q_1\|_A Q_2 \rnar{\alpha}{} Q'_1\|_A Q'_2}\\
\name{Rel$_{r}$} \sos{Q\rnar{\alpha}{}Q'}
{Q[\Phi]\rnar{\Phi(\alpha)}{}Q'[\Phi]}\quad \quad
\name{Rec$_{r}$} \sos{Q \{\rec x. Q/x\} \rnar{\alpha}{} Q'}
{\rec x.Q \rnar{\alpha}{}  Q'}\\
\end{array}\]
\caption{Reading Behaviour of PAFAS$_r$
processes}\label{ReadBehaviour}
\end{table}
Most of the rules in Table~\ref{ReadBehaviour} say that the execution
of reading actions does not change the state of a term $Q$. Rule
\name{Read$_{r2}$} is crucial to manage arbitrarily nested reading
actions; contrast it with \name{Read$_{o}$}. Due to
technical reasons, rule \name{Rec$_r$} allows unfolding of recursive
terms; thus e.g. $\rec x. \,a \rop b .x \rnar{a}{} a \rop b. ( \rec x.
\, a \rop b .x )$. Notice that this leads to a timed bisimilar
process, cf.~Section~\ref{secmapping}.

To give SOS-rules for the time steps of process terms, we consider
(partial) {\it time-steps} like $Q \nar{X}{r} Q'$ where
the set $X \subseteq \B$ (called a {\em refusal set})
consists of non-urgent actions. Hence $Q$ is justified in delaying,
i.e.\ refusing them; $Q$ can take part in a real time step only if it
has to synchronise on its urgent actions, and these are delayed by the
environment. If $X = \B$ then $Q$ is fully justified in performing
this full unit-time step; i.e., $Q$ can perform it independently of
the environment. If $Q \nar{\B}{r} Q'$, we write $Q \nar{1}{}Q'$; we
say that $Q$ performs a {\it 1-step}.

\begin{definition}\label{PAFASRT}\rm ({\sl refusal transitional
semantics})
The inference rules in Table~\ref{rt-semantics} define $\nar{X}{r}
\subseteq \Pg \times \Pg$ where $X \subseteq \B$. A refusal trace of a
term $Q \in \Pg$ records from a {\em run} of $Q$ which visible actions
are performed ($Q \nar{a}{} Q'$, $a \in \B$) and which actions $Q$
refuses to perform when time elapses ($Q\nar{X}{r} Q'$, $X \subseteq
\B$); i.e.\ a  refusal trace of $Q$ is the sequence of actions from
$\B$ and refusal sets $\subseteq \B$ occurring in a finite transition
sequence from $Q$ (abstracting from $\tau$-transitions).
\end{definition}

\begin{table}[tbh]
\small
\[
\begin{array}{c}
\name{Nil$_{t}$}\; \sos{}{\Nil \nar{X}{r} \Nil}\qquad\qquad

\name{Pref$_{t1}$}\; \sos{}{\alpha.P \nar{X}{r}
\underline{\alpha}.P}\qquad\qquad

\name{Pref$_{t2}$}\; \sos{\alpha \notin X \cup \{\tau\}}
{\underline{\alpha}.P \nar{X}{r} \underline{\alpha}.P}\\

\name{Read$_{t1}$}\; \sos{Q \nar{X}{r} Q'}{\alpha \rop Q \nar{X}{r}
\underline{\alpha} \rop Q'}\qquad \qquad

\name{Read$_{t2}$}\; \sos{Q \nar{X}{r} Q', \; \alpha \notin X \cup
\{\tau\}}
{\underline{\alpha} \rop Q \nar{X}{r} \underline{\alpha} \rop Q'}\\

\name{Sum$_{t}$}\; \sos{Q_i\nar{X}{r}Q'_i \mbox{ for } i=1,2}{Q_1+Q_2
\nar{X}{r} Q'_1 + Q'_2} \qquad \qquad

\name{Rel$_{t}$} \sos{Q\nar{\Phi^{-1}(X \cup \{\tau\})\backslash
\{\tau\}}{r}Q'} {Q[\Phi]\nar{X}{r}Q'[\Phi]}\\ 

\name{Rec$_{t}$} \sos{Q \{\rec x. Q/x\} \nar{X}{r} Q'}
{\rec x.Q \nar{X}{r} Q'}\qquad

\name{Par$_{t}$} \sos{Q_i \nar{X_i}{r} Q'_i \mbox{ for } i=1,2, X
\subseteq (A \cap (X_1 \cup X_2)) \cup ((X_1 \cap X_2)\backslash A)}
{Q_1\|_A Q_2 \nar{X}{r} Q'_1\|_A Q'_2}\\
\end{array}\]
\caption{Refusal transitional semantics of PAFAS$_r$ processes}
\label{rt-semantics}
\end{table}

Rule \name{Pref$_{t_1}$} says that $\alpha.P$ can let time pass and
refuse to perform any action while rule \name{Pref$_{t_2}$} says that
$\underline{\alpha}.P$ can let time pass in an appropriate context,
but cannot refuse the action $\alpha$. Process $\underline{\tau}.P$
cannot let time pass at all since, in any context,
$\underline{\tau}.P$ has to perform $\tau$ before time can pass
further. Rule~\name{Par$_{t}$} defines which actions a parallel
composition can refuse during a time-step. $Q_1 \|_A Q_2$ can refuse
the action $\alpha$ if either $\alpha \notin A$ and $\alpha$ can be
refused by both $Q_1$ and $Q_2$ or $\alpha \in A$ and at least one of
$Q_1$ and $Q_2$ can delay it, forcing the other $Q_i$ to wait. Thus,
an action is urgent (cannot be further delayed) only when all
synchronising \lq local\rq\ actions are urgent. The other rules are as
expected.

\begin{example}\rm\label{exboolarray}
As an example for the definitions given so far, consider
an {\em array} with two Boolean values $t$ and $f$ and define
its behaviour as $B_{tf} \equiv P_t \,\|_{A} Q_f$ where  $P_t
\Leftarrow r_{tt} \rop (r^1_t \rop w^1_f.P_f) + r_{tf} \rop (r^1_t
\rop w^1_f.P_f)$, $Q_f \Leftarrow r_{tf} \rop (r^2_f \rop w^2_t. Q_t)
+ r_{ff} \rop (r^2_f \rop w^2_t.Q_t)$ and $A=\{r_{ij} \,|\, i,j \in
\{t,f\} \} $. Actions $r_{ij}$, where $i,j \in \{t,f\}$, allow reading
both entries at the same time, while $r^k_j$ and $w^k_j$ represent,
resp., the reading and the writing of the value $j \in \{t,f\}$ for
the entry $k \in \{1,2\}$. By rules~\name{Read$_{r1}$}
and~\name{Read$_{r2}$},  $B_{tf} \rnar{r_{tf}}{} B_{tf}$ and $B_{tf}
\rnar{r^1_{t}}{} B_{tf}$ describing non-blocking reading. $P_t$ offers
a choice between $r_{tf}$ and $r_{tt}$, where synchronisation
disallows the latter. Performing $w^1_f$ after a 1-step does not
change the second component, so $r^2_f$ is still urgent; this shows
that  $w^1_f$ does not block $r^2_f$. With just one type of action
transition, $P_{t}$ would lose the prefix $r_{tf} \rop$ when
performing $r^1_{t}$. Only the execution of an ordinary action can
change the state of the array, e.g.\ $B_{tf} \onar{w^1_{f}}{} B_{ff}
\equiv P_f \,\|_{A} Q_f$ by Rule~\name{Read$_o$}.
\end{example}
In \cite{CVJ02}, it is shown that inclusion of refusal traces
characterises an efficiency preorder which is intuitively justified by
a testing scenario. In this sense, e.g.\ $P \equiv a \rop b$ is faster
than the functionally equivalent $Q \equiv \rec x.\, (a.x +b)$, since
only the latter has the refusal traces $1a(1a)^{\ast}$: after $1a$,
$Q$ returns to itself, since recursion unfolding creates fresh $a$ and
$b$; intuitively, $b$ is disabled during the occurrence of $a$, so $a$
and also $b$ can be delayed again. In contrast, after a time step and
any number of $a$s, $P$ turns into $\underline{a} \rop \underline{b}$
and no further 1-step is possible. Since read actions do not block or
delay other activities, they make processes faster and, hence, have
an impact on timed behaviour of systems. If $a$ models the reading of
a value stored by $P$ or $Q$ and two parallel processes want to read
it, this should take at most time 1 in a setting with non-blocking
reads. And indeed, whereas $Q \,\|_{\{a\}}\, (a \,\|_{\emptyset}\, a)$
has the refusal trace $1a1a$, this behaviour is not possible for $P
\,\|_{\{a\}}\, (a\,\|_{\emptyset}\, a)$. Thus, $P$ offers a
\emph{faster} service.

Another application of refusal traces is the modelling of {\it weak
fairness of actions}. Weak fairness requires that an action must be
performed whenever continuously enabled in a run. Thus, a run from $P$
with infinitely many $a$'s is not fair; the read action does not block
$b$ or change the state, so the same $b$ is  always enabled but never
performed. In contrast, if $Q$ performs $a$, a fresh $b$ is created;
in conformance to~\cite{CS87}, a run with infinitely many $a$'s is
fair. In~\cite{CDV09-2}, generalising \cite{CDV06}, fair traces for
PAFAS$_r$ (and PAFAS$_s$) are first defined in an intuitive, but very
complex fashion in the spirit of~\cite{CS87} and then characterised:
they are the sequences of visible actions occurring in transition
sequences with infinitely many 1-steps\footnote{Observe
that~\cite{CDV09} just contains the application presented
in~\cite{CDV09-2}; PAFAS$_r$ is not treated there at all.}. Due to
lack of space, we cannot properly formulate this as a theorem, but
take it as a (time-based) {\bf definition} of \emph{fair traces}
instead; ${\sf FairL}(R)$ is the set of fair traces of $R$. With this,
infinitely many $a$'s are a fair trace of $Q$ since it can repeat $1a$
indefinitely, but the fair traces of finite-state $P$ are those that
end with $b$. This shows an added expressivity of read prefixes:

\begin{theorem}\label{theorem-expr-read-actions}
If $R\in \Pg$ is a finite-state process without read-prefixes and with
sort $\synsort{R}=\{a,b\}$, then ${\sf FairL}(R) \neq \{a^i b \,|\, i
\geq 0\} = {\sf FairL}(a \rop b)$.
\end{theorem}

We can view fairness as imposing a kind of priority for $b$ in $P$
since, in contrast  to $a$, it must be executed in a fair trace.
This is of course very different from the usual treatment of
priorities \cite{HPA2001}, since $a$ can be prefered to $b$ for a
number of times. 
The following example shows that read actions can model more than two
levels of priority.
\begin{example}\rm\label{remark:priority1}
In $P \equiv a \rop ( (\rec x. b.x) \;\|_{\{b\}}\; b \rop c)$, there
are three levels of priority: in a fair trace we can perform
arbitrarily many $a$'s while both $b$ and $c$ remain enabled and have
priority -- so far, we can have at most one 1-step. If $b$ occurs, the
action $a$ disappears but we can perform arbitrarily many $b$'s while
$c$ remains enabled and has priority -- with, still, at most one
1-step. Formally, with a 1-step $P$ evolves into $\underline{P} \equiv
\underline{a} \rop (\underline{b}. (\rec x. b.x) \;\|_{\{b\}}\;
\underline{b} \rop \underline{c})$. $\underline{P}$ can perform an $a$
to itself, a $c$ (and become $\underline{b}.(\rec x.b.x)
\;\|_{\{b\}}\;\Nil$), or  repeated $b$'s to $((\rec x. b.x)
\;\|_{\{b\}}\; \underline{b} \rop \underline{c}$; no further  1-steps
are possible due to the urgent $c$; so in a fair trace, finally $c$ is
performed to $((\rec x. b.x) \;\|_{\{b\}}\; \Nil)$ -- where infinitely
many 1-steps are possible.
\end{example}

\section{A read operator with a simpler semantics}\label{PAFAS-s}
The special reading transitions of PAFAS$_r$ are needed to properly derive
e.g.\ $ P \equiv a \rop b \rop Q \nar{b}{} a \rop b\rop Q$. To get a
simpler semantics, the idea is to collect all enabled reading actions of a
\lq sequential component\rq\ in a set and write e.g.\ $P$ as $\{a, b\}\rop
c$. Thus, we define a new kind of read operator $\{\mu_1, \ldots, \mu_n\}
\rop Q$ with a slightly different syntax. In this way we try to avoid terms
with nested reading actions and, as a consequence, we can describe the
behaviour of the new PAFAS$_s$ processes by means of a simpler timed
operational semantics with  just one type of action transitions. A price to
pay is that not all PAFAS$_s$ processes have a reasonable semantics; but
the subset with a reasonable semantics is practically expressive
enough (e.g. for expressing MUTEX solutions adequately)
due to the {\em set} of reading actions, cf.~\cite{CDV09}.

The sets $\Sg_1$ of {\em initial (timed) process terms} $P$ and $\Sg$ of
(general) {\em (timed) process terms} $Q$ is generated by the following
grammar:

\begin{eqnarray*}
P&::=& \Nil \gor x \gor \alpha. P \gor \{\alpha_1, \ldots, \alpha_n\}
\rop P \gor P+P \gor P \:\|_A \: P \gor P[\Phi] \gor \rec x.P\\ Q
&::=& P \gor \underline{\alpha} .P \gor \{\mu_1, \ldots, \mu_n\} \rop
Q \gor Q+Q \gor Q \:\|_A\: Q \gor Q[\Phi] \gor \rec x.Q
\end{eqnarray*}
\noindent where $\Nil$ is a constant, $x \in \X$, $\alpha\in\Bt$,
$\{\alpha_1, \ldots, \alpha_n \}\subseteq \Bt$ finite, $\{\mu_1,
\ldots, \mu_n\}$ is a finite subset of $\Bt \cup \underline{\B}_\tau$
that cannot contain two copies (one lazy and the other one urgent) of
the same action $\alpha$, i.e.  $\big| \{\alpha, \underline{\alpha}\}
\cap \{\mu_1, \ldots, \mu_n\}\big| \leq 1$ \ for any $\alpha\in \Bt$.
Again, $\Phi$ is a general relabelling function and $A \subseteq \B$
possibly infinite. Also in this section, recursion is guarded. The
sets of closed timed process terms in $\Sg$ and $\Sg_1$, simply called
{\em processes} and {\em initial processes} resp., are $\SG$ and
$\SG_1$ resp. 

\begin{definition}\label{PsACTSOS}\rm({\sl functional operational
semantics}) The SOS-rules defining the transition relations $\nar{\alpha}{}
\subseteq(\Sg\times\Sg)$ (the {\em action transitions}) are those in
Table~\ref{OrdBehaviour}\footnote{To be formally precise: we have to
replace all arrows $\onar{}{}$ in Table~\ref{OrdBehaviour} by $\nar{}{}$.}
where we replace the rule \name{Read$_o$} with:
\small
$$\begin{array}{c}
\name{Read$_{s1}$}\; \sos{ \mu_i \in \{\alpha, \underline{\alpha}\}}
{\{\mu_1, \ldots, \mu_n\} \rop Q \nar{\alpha}{} \{\mu_1, \ldots,
\mu_n\} \rop Q}\qquad
\name{Read$_{s2}$}\; \sos{Q \nar{\alpha}{} Q'}
{\{\mu_1, \ldots, \mu_n\} \rop Q \nar{\alpha}{} Q'}
\end{array}$$
\end{definition}

\begin{definition}\label{PAFASsRT}\rm ({\sl refusal transitional
semantics}) The inference rules defining the transition relation
$\nar{X}{r} \subseteq \Sg \times \Sg$ where $X \subseteq \B$ are those in
Table~\ref{rt-semantics} where we replace the rules~\name{Read$_{t1}$}
and~\name{Read$_{t2}$} with:
\small
$$\begin{array}{c}
\name{Read$_{t}$}\; \sos{Q \nar{X}{r} Q', \; \ur(\{\mu_1, \ldots,
\mu_n\}) \cap (X \cup \{\tau\}) = \emptyset}
{\{\mu_1, \ldots, \mu_n\} \rop Q \nar{X}{r}
\underline{\{\mu_1, \ldots, \mu_n\}} \rop Q'}
\end{array}$$
\normalsize
\noindent where $\ur(\{\mu_1, \ldots, \mu_n\}) = \{\alpha \,|\, \mu_i
= \underline{\alpha} \mbox{ for some } i \in [1,n]\}$ and
$\underline{\{\mu_1, \ldots, \mu_n\}}$ is the set obtained from
$\{\mu_1, \ldots, \mu_n\}$ by replacing each $\alpha$ by
$\underline{\alpha}$.
\end{definition}
A term $Q\in \Sg$ is {\em read-guarded} if every subterm of $Q$ of the
form $\{\mu_1, \ldots, \mu_n\} \rop Q'$ is in the scope of some action
prefix $\mu.()$. $Q\in \Sg$ is {\em read-proper}
if each subterm $Q_1 + Q_2$ is read-guarded and, for each subterm
$\{\mu_1, \ldots, \mu_n\} \rop Q_1$, $Q_1$ is read-guarded.
We say that $Q$ is {\em $x$-proper} if any free $x$ is guarded in any
subterm $Q_1 + Q_2$, $\{\mu_1, \cdots, \mu_n\} \rop Q_1$ and $\rec y.
Q_1$. $Q$ is {\em rec-proper} if for any subterm $\rec x.Q_1$, $Q_1$
is either read-guarded or $x$-proper. A term $Q$ is {\em proper} if it
is read- and rec-proper. Below, we will prove that proper terms have a
reasonable semantics by relating them to PAFAS$_r$ processes with the
same behaviour. An important feature of properness is that processes
without read-prefixes are proper.

According to the definitions given so far, neither $P\equiv \{a\} \rop
\{b\} \rop Q$ nor $P'\equiv \{a\}\rop Q' + \{b\} \rop Q $ are
read-proper because of $\{b\} \rop Q$. An essential idea of reading is
that it does not change the state of a process and therefore does not
block other actions. With this, we should have $P \nar{b}{} P$, but
really we have $P \nar{b}{} \{b\} \rop Q$. Similarly, we have
$P'\nar{b}{} \{b\} \rop Q$ instead of $P' \nar{b}{} P'$. Hence, we
exclude such processes. There is also a problem with the term $P
\equiv \rec x. \{a\} \rop b.(c + x)$. Indeed, $P$ can perform a $b$
and  evolve to $c + \rec x. \{a\} \rop b.(c + x)$ which is not
read-proper. Since the body of this recursion is not read-guarded, $x$
has to be treated as a read-prefix term, i.e.\ the body has to be
$x$-proper. A subtle detail is the consideration of recursive subterms
in the definition of $x$-proper. Without this detail,  $Q
\equiv \rec x. \{a\} \rop b. \rec y. (c.(c+y) \,\|_\emptyset\, x)$
would be proper. But, $Q \nar{b}{} \rec y. (c.(c + y)
\,\|_\emptyset\, Q)   \nar{c}{} ( c +  \rec y. (c.(c + y)
\,\|_\emptyset\, Q)) \,\|_\emptyset\, Q$. Notice that  $ \rec y.
(c.(c + y)  \,\|_\emptyset\, Q)$, and hence $ c +  \rec y.
(c.(c + y)  \,\|_\emptyset\, Q)$, is not read-proper.

In contrast to the restriction to proper terms, we can freely use
read-prefixes in PAFAS$_r$, see e.g.\ the process in
Example~\ref{exboolarray}; this would have the {\em wrong semantics}
in PAFAS$_s$, i.e.\ if we change $r_{ij} \rop {}$ and $r^k_j \rop{}$
(for $i,j \in \{t,f\}$ and $k \in \{1,2\}$) into $\{r_{ij}\} \rop$ and
$\{r^k_j\} \rop$. The restriction only makes sense because of
Prop.~\ref{propRPclosure}, which requires a careful, detailed proof.
\begin{proposition}\label{propRPclosure}
Let $Q\in\Sg$ be proper.  $Q\nar{\alpha}{}Q'$ or $Q\nar{X}{r}Q'$
implies $Q'$ proper. \end{proposition}
Actually, the result in~\cite{CDV09-2} is not correct since we used an
insufficient restriction there. But, luckily the PAFAS$_s$ process we
used to model Dekker's MUTEX algorithm is proper. This can been easily
seen since proper processes are closed w.r.t. parallel composition and
relabelling. %

\section{Expressivity of PAFAS$_s$}\label{secmapping}
In this section we compare the expressivity of PAFAS$_s$ with that of
PAFAS$_r$ and Petri nets. A first result shows that for each proper
$Q\in\Sg$ there is a term in $\Pg$ whose behaviour is (timed)
bisimilar and even isomorphic to that of $Q$.

\begin{definition}\rm\label{def-bisimulation}({\sl timed bisimulation})
A binary relation ${\cal S} \subseteq \PG \times \PG$ over processes
is a {\em timed bisimulation} if $(Q,R) \in {\cal S}$ implies, for all
$\alpha \in \Bt$ and all $X \subset \B$:

\begin{itemize}
\item [-]
whenever  $Q \rnar{\alpha}{} Q'$ ($Q \onar{\alpha}{} Q'$, $Q
\nar{X}{r} Q'$) then, for some $R'$, $R \rnar{\alpha}{} R'$ ($R
\onar{\alpha}{} R'$, $R \nar{X}{r} R'$, resp.) and $(Q',R') \in {\cal
S}$;
\item [-] whenever  $R \rnar{\alpha}{} R'$ ($R \onar{\alpha}{} R'$, $R
\nar{X}{r} R'$ ) then, for some $Q'$, $Q \rnar{\alpha}{} Q'$ ($Q
\onar{\alpha}{} Q'$, $Q \nar{X}{r} Q'$, resp.) and $(Q',R') \in {\cal
S}$.
\end{itemize}

\noindent
Two processes $Q, R \in \Pg$ are timed bisimilar ({\em bisimilar} for
short, written $Q \bisimilar R$) if $(Q,R) \in {\cal S}$ for some
timed bisimulation ${\cal S}$. This definition is extended to open
terms as usual; two open terms are bisimilar if they are so for all
closed substitutions. It can be proved in a standard fashion that
timed bisimilarity is a {\em congruence} w.r.t. all PAFAS$_r$
operators. The same definition, but omitting the reading transitions,
applies to PAFAS$_s$.
\end{definition}
We start by providing a translation function $\denoter{\_}$ that maps
terms in $\Sg$ into corresponding terms in $\Pg$; to regard
$\denoter{\_}$ as a function in the read-case, we have to assume that
actions are totally ordered, and that the actions of a read-set are
listed according to this order.

\begin{definition}\label{defdenote}\rm({\sl a translation function})
For $Q\in\Sg$ proper, $\denoter{Q}$ is defined by induction on $Q$
(subterms of $Q$ are also proper) as follows :
\begin{tabbing}
\qquad Nil, Var, Pref : \quad \=
$\denoter{\Nil} \equiv \Nil$, \qquad $\denoter{x} \equiv x$, \qquad
$\denoter{\mu.P} \equiv \mu.\denoter{P}$\\
\qquad Read: \> $\denoter{\{\mu_1, \ldots, \mu_n\} \rop Q} \equiv
\mu_1 \rop \ldots \rop \mu_n \rop \denoter{Q}$\\
\qquad Sum: \>
$\denoter{Q_1 + Q_2} \equiv \denoter{Q_1} + \denoter{Q_2}$ \\
\qquad Par: \>
$\denoter{Q_1 \:\|_A\: Q_2} \equiv \denoter{Q_1} \:\|_A\:
\denoter{Q_2}$ \\ %
\qquad Rel: \>
$\denoter{Q[\Phi]} \equiv \denoter{Q}[\Phi]$\\
\qquad Rec: \> $\denoter{\rec x.Q} \equiv \rec x. \denoter{Q}$
\end{tabbing}
\end{definition}
This translation is pretty obvious, but its correctness
is not; observe that Theorem~\ref{teo:isomorphism} does not hold for
all PAFAS$_s$ processes; cf.\ the processes $P \equiv \{a\} \rop
\{b\} \rop Q$ and $P' \equiv \{a\} \rop Q' + \{b\} \rop Q$ at the end
of Section~\ref{PAFAS-s}.
Function $\denoter{}$ is injective on proper terms; except for the
read-case, this is easy since $\denoter{}$ preserves all other
operators. In the read-case, $Q$ is read-guarded, i.e.\ the
top-operator of $Q$ and $\denoter{Q}$ is not $\rop$;  the read-set can
be read off from $\denoter{\{\mu_1, \ldots, \mu_n\} \rop Q}$ as the
maximal sequence of $\rop$-prefixes the term starts with. With this
observation, the following result, together with
Prop.~\ref{propRPclosure}, shows that $\denoter{}$ is an isomorphism
between labelled transition systems, if we restrict them, on the one
hand, to proper terms and their transitions and, on the other, to the
images of proper terms and the transitions of these images.
\begin{theorem}\label{teo:isomorphism}\rm
For all proper $Q\in\Sg$:
\begin{enumerate}
\item $Q\nar{\alpha}{}Q'$ ($Q\nar{X}{r}Q'$) implies $\denoter{Q}
\nar{\alpha}{}\denoter{Q'}$ ($\denoter{Q}\nar{X}{r}\denoter{Q'}$,
resp.);

\item if $\denoter{Q}\nar{\alpha}{}Q''$ ($\denoter{Q}\nar{X}{r}Q''$)
then $Q\nar{\alpha}{}Q'$ ($Q\nar{X}{r}Q'$) with $\denoter{Q'} \equiv
Q''$.
\end{enumerate}
\end{theorem}

The above theorem shows that the expressivity of proper PAFAS$_s$ processes
is at most that of PAFAS$_r$. On the other hand, it is enough to model safe
Petri nets with read-arcs. To illustrate the proof idea, which is based on
a well-known view of a net as a parallel composition, consider an empty
place of a net with preset $\{t_1, t_2\}$ and postset $\{t_3, t_4\}$, and
being read by $t_5$ and $t_6$. This is translated into process $P_0$ with
$P_0 \Leftarrow t_1.P_1 + t_2.P_1$ and $P_1 \Leftarrow \{t_5, t_6\} \rop
(t_3.P_0 + t_4.P_0)$; $P_1$ models the marked place. All the analogous
translations of places are composed in parallel, synchronising each time
over all common actions (e.g.\ net transitions). Finally, a relabelling
corresponding to the labelling of the net is applied.

\begin{theorem}
For each safe Petri nets with read-arcs in~\cite{VOGLER02} there is a
bisimilar proper PAFAS$_s$ process.
\end{theorem}
\section{The backward translation from PAFAS$_r$ to
PAFAS$_s$}\label{sec-backward-mapping}
In this section we study the problem whether PAFAS$_r$ is more
expressive than PAFAS$_s$ or whether each PAFAS$_r$ term can be
translated into a bisimilar proper PAFAS$_s$ term. We first exhibit a
subset of $\Pg$ that is essentially the image of $\denoter{.}$ and so
has an easy translation; we say these terms are in {\em read normal
form (RNF)} (see Def.~\ref{def-readnf}). We then discuss how PAFAS$_r$
terms can be brought into RNF and illustrate, by means of examples,
the problems of such a normalisation.
\begin{definition}\label{def-readnf}\rm ({\sl read normal form})
For PAFAS$_r$ terms, we define read-guarded, and $x$- and rec-proper
as above except for considering read-action prefixes instead of
read-set prefixes. We call such a term {\em ra-proper} if each subterm
$Q_1 + Q_2$ is read-guarded, and for each subterm $\mu \rop Q'$ either
 $Q'$ is read-guarded or $Q' \equiv \nu \rop Q''$.  A term is {\em
RNF} if it is rec- and ra-proper. The sets of terms and processes in
RNF are denoted by $\Nf$ and $\NF$, resp.
\end{definition}

Below we provide the function that translates each $Q \in \Nf$ into a
proper term in $\Sg$. We will need an additional function to deal with
read prefixes. A term such as $\mu_1 \rop Q$ is in RNF if either $Q$
is
read-guarded or, by iterative applications of Def.~\ref{def-readnf},
$Q$ has the form $\mu_1 \rop \cdots \rop \mu_n \rop Q_n$ where $Q_n
\in \Nf$ is read-guarded. In the latter case, the actions $\mu_1,
\cdots, \mu_n$ must be collected in a read set.
Since read sets cannot contain multiple copies (lazy and
urgent) of the same action $\alpha$, we use the following
notation: if $\mu_1, \cdots, \mu_n$ are generic actions in $\Bt \cup
\underline{\B}_\tau$, $\denote{\mu_1, \cdots, \mu_n}$ denotes the set
of actions $\{\nu_1, \cdots, \nu_m \}$ such that: 
$\exists \, i \in [1,m]$ with $\nu_i = \underline{\alpha}$ iff
$\exists \, j \in [1,n]$ with $\mu_j = \underline{\alpha}$; 
(2) $\exists \, i \in [1,m]$ with $\nu_i = \alpha$ iff $\exists \, j
\in [1,n]$ such that $\mu_j =\alpha$ and, for each $k \in [1,n]$,
$\mu_k \neq \underline{\alpha}$. 
\begin{definition}\label{def-mapping-r2s}\rm({\sl a translation
function from $\Nf$ to $\Sg$}) For $Q \in \Nf$, we define the process
term $\denotes{Q} \in \Sg$ by induction on  $Q$ as in
Definition~\ref{defdenote} except for:
%
\begin{tabbing}
Read: $\denotes{\mu_1 \rop Q} \equiv$ \=
$\denote{\mu_1, \cdots, \mu_n} \rop \denotes{Q_n}$ \\
\>  if  $Q \equiv  \mu_1 \rop \cdots
\rop \mu_n \rop  Q_n  \mbox{ and } Q_n \mbox{ is read-guarded}
$
\end{tabbing}
%
\end{definition}
With the laws L1 and L2 below, we can rearrange successive read-action
prefixes in a process in RNF such that the result is in the  image of
$\denoter{}$, which essentially proves the second item of following
result.

\begin{theorem}\label{teo-inverse-mapping}\rm
For all $Q\in\Nf$:
\begin{enumerate}
\item $Q \nar{\alpha}{} Q'$ or $Q \nar{X}{r} Q'$ imply $Q' \in \Nf$;

\item $Q$ and $\denotes{Q}$ are timed bisimilar (in the sense of
PAFAS$_s$).
\end{enumerate}
\end{theorem}
\subsection*{Translating terms into read normal
form}\label{sec-not-nf}
For translating a term that is {\em not} in read normal form, one idea
is to use laws to rewrite the term into a bisimilar one in RNF. E.g.\,
although $(a \rop b) + c $ does not belong to $\Nf$, it has the same
timed behaviour as $ a \rop (b+c) \in \Nf$, cf.\ L3.

Besides commutativity and associativity of $+$ and $\|$, we have shown
the laws in Fig.~\ref{fig-axioms}. Here, $\Phi\{a \rightarrow
\alpha\}$  denotes the relabelling function that renames $a$ to
$\alpha$, and all other actions as $\Phi$. For the discussion, we also
write $[a \rightarrow \alpha]$ as a shorthand for $\Phi_{I}\{a
\rightarrow \alpha\}$ where $\Phi_I$ is the identity relabelling
function.
\begin{figure}[ht]
\begin{center}
\begin{tabular}{| l | l  |}
\hline
L1 &  $\mu \rop (\nu \rop Q) \sim \nu \rop (\mu \rop Q)$ \\
L2 & $\alpha \rop (\mu \rop Q) \sim \mu \rop Q$,
$\underline{\alpha} \rop (\mu \rop Q) \sim \underline{\alpha} \rop Q$
provided that $\mu \in \{\alpha, \underline{\alpha}\}$\\
L3 & $(\mu \rop Q) + R \sim  \mu \rop (Q + R)$ \\
L4 & $a \rop (Q_1 \;\|_A\; Q_2) \sim ((a \rop Q_1) \;\|_{A \:\cup\:
\{a\}}\; (a \rop Q_2))$,\\ & $\underline{a} \rop (Q_1 \;\|_A\; Q_2)
\sim ((\underline{a} \rop Q_1) \;\|_{A \:\cup\:
\{a\}}\;(\underline{a} \rop Q_2))$ provided that $a \notin
\synsort{Q}$ \\
L5 & $(\alpha \rop Q)[\Phi] \sim \Phi(\alpha) \rop (Q[\Phi])$,
$(\underline{\alpha} \rop Q)[\Phi] \sim \underline{\Phi(\alpha)} \rop
(Q[\Phi])$\\
L6 & $(Q[\Phi])[\Psi] \sim Q [\Psi \circ \Phi]$\\
L7& $\rec x. Q \sim Q \{\rec x.Q /x\}$\\
\hline
\end{tabular}
\end{center}
\caption{A set of laws}\label{fig-axioms}
\end{figure}
The idea of the translation into RNF is to perform rewriting by
induction on the term size; since action-prefix, parallel composition
and relabelling preserve RNF, these operators are no problem.
Read-prefixes $\mu \rop Q$ can be dealt with distributing $\mu$ among
$Q$'s components. But choice and recursion pose still unsolved
problems.

Regarding read prefixes, we have to show the stronger claim that for
each $Q$ in RNF we can normalise $\mu \rop Q$ in such a way that, for
any variable $y$, $y$ guarded in $Q$ implies $y$ guarded in the RNF,
and if additionally $Q$ is $y$-proper this is also preserved. The
proof is by induction on $Q$; some cases are easy because $\mu \rop Q$
is in RNF itself (by the definition of RNF or by induction). We
consider one of the three remaining cases, namely the Par-case. The
Rel-case is easier, while the Rec-case is much more complicated. Their
proofs can be found in the appendix. For a fresh action $a$ we have:

$$\alpha \rop (Q_1 \,\|_A\, Q_2 ) \sim (a \rop (Q_1 \,\|_A\, Q_2
))[a \to \alpha] \sim ((a \rop Q_1) \,\|_{A \cup \{a\} }\, (a \rop
Q_2))[a \to \alpha]$$

\noindent by L4, and then we are done by induction. The case of an
$\underline{\alpha}$-read-prefix is similar.

The case of choice is particularly tricky whenever one of the two
alternatives  is a parallel composition. Hence, we now concentrate on
the following problem: {\em let $Q, R \equiv R_1 \,\|_A\, R_2$ be
terms in RNF; is there an $S$ in RNF such that $S \sim Q + R$}?

First, observe that we can rewrite $Q$ into $Q'$ by replacing all
actions (also in relabellings) by fresh copies, such that $Q'$ and $R$
have disjoint sorts. Then, we can try to bring $Q'+R$ into RNF and
finally apply a relabelling that `undoes'  the rewrite (cf.\ the last
example above). This would give us a bisimilar term in RNF for $Q+ R$.
From now on we assume that $Q$ and $R$ have disjoint sorts.

If $Q$ is {\em deterministic} (i.e.\ it never performs $\tau$ and
never performs an action in two different ways), we have the law
$Q + (R_1 \;\|_A\; R_2) \sim (Q + R_1) \;\|_{A \, \cup\,
\synsort{Q}}\; (Q + R_2)$. Thus, to find $S$ we now simply normalise
the two components inductively.
In general, this law fails: for $Q \equiv a.b + a.c$, $Q + R$ evolves
with $a$ into either $b$ or $c$. But $(Q + R_1) \;\|_{A \, \cup\,
\{a,b,c\}}\; (Q + R_2)$ can perform $a$ and evolve into the deadlocked
$b \;\|_{A \, \cup\, \{a,b,c\}}\; c$. A new idea that will work in
many cases is to replace the second copy of $Q$ by its `top-part' that
can perform the same time steps and the same initial actions as $Q$,
but deadlocks after an ordinary action; additionally, not all of
$\synsort Q$ but only the initial actions are added to the
synchronisation set: in our example, $((a.b + a.c) + R_1) \;\|_{A \,
\cup\, \{a\}}\; (a + R_2)$ is  bisimilar to $Q+R$ and could, in
principle, be normalised inductively.
This idea must be adapted in case of read prefixes. Consider $Q\equiv
a \rop b.c$; here, the top-part is $a \rop b$, i.e.\ $Q+R$ is
bisimilar to $(a \rop b.c + R_1) \;\|_{A \, \cup\, \{a,b\}}\; (a \rop
b + R_2)$ (in particular, both terms remain unchanged when performing
$a$). Another problem is that initial actions  may also be performed
later, e.g.\ in $Q\equiv a \rop b.a$; again, rewriting plus later
relabelling helps. In the example, $Q+R$ is bisimilar to $((e \rop b.c
+ R_1) \;\|_{A \, \cup\, \{e,b\}}\; (e \rop b + R_2)) [e \rightarrow
a]$, and the terms $e \rop b.c + R_1$ and $e \rop b + R_2$ are again
smaller than $Q+R$.

But what is the top-part for $Q\equiv a \;\|_{\emptyset}\; b$? Action
$a$ can be performed initially, but also after $b$. If we could
transform $Q$ into $a.b+b.a$, the top-part would be $a+b$, and using
rewriting plus later relabelling solves the problem. But unfortunately
$Q\sim a.b+b.a$ is wrong: when performing $1a$,  these terms end up in
$\Nil \;\|_{\emptyset}\;\underline{b}$ and $b$ resp., which are not
timed bisimilar due to partial time step $\{b\}$.

Finding the top-part of parallel compositions seems to be related to
finding a suitable expansion law. But even for standard PAFAS, such a
law is not known. Thus, our general proof idea does not work so far,
due to problems with choice terms. Also the treatment of recursion is
not clear yet; an expansion law would certainly help.
At least, we have identified a fragment of PAFAS$_r$ which does not
have additional expressivity.
\begin{theorem}
If all choice and recursive subterms of a PAFAS$_r$ process are in RNF
then there is a bisimilar PAFAS$_s$ process.
\end{theorem}

\section{Conclusions and Future Work}\label{sec-conclusion}
We have studied two different ways to enhance PAFAS with non-blocking
reading actions. We have first added reading in the form of a
read-action prefix operator and proved that this adds expressivity
w.r.t.\ fair behaviour. This operator is very flexible, but has a
slightly complex semantics. To reduce complexity, we have introduced a
read-set prefix operator with a simpler semantics, but with syntactic
restrictions. For the second operator, it is not immediately clear
whether its operational semantics models reading behaviour adequately.
We could prove this by translating proper PAFAS$_s$ terms into
PAFAS$_r$ terms with the same timed behaviour. We also show that
PAFAS$_s$ is strong enough to model Petri nets with read-arcs.

It is still not clear whether PAFAS$_r$ is more expressive than the
restricted PAFAS$_s$. We presented some ideas how a respective
translation could work; these are based on some algebraic laws that
are also interesting in their own right. In the future we will try to
complete this translation. This is related to finding an expansion law
for generic PAFAS$_r$ (and PAFAS) terms. Such an expansion law should
also provide us with an axiomatisation for the full PAFAS language.
Some results can be found in~\cite{VJ2000} where a fragment of the
language that just consists of prefix and choice has been
axiomatised.

We plan to use read prefixes for modelling systems and comparing their
efficiency or proving them correct under the progress assumption. A
first correctness proof (for Dekker's MUTEX algorithm) with the aid of
the automated verification tool FASE has been presented
in~\cite{CDV09}. %

\nocite{*}
\bibliographystyle{eptcs}

\end{document}